\documentclass[a4paper]{jpconf}
\usepackage{graphicx}
\begin{document}
\title{Measurement of the fraction of top quark pair events produced via gluon-gluon fusion at the Tevatron in lepton+jets final states}

\author{Sungwoong Cho$^a$,
    Suyong Choi$^a$,
    Sehwook Lee$^a$,
    JaeHoon Lim$^a$, and
    SungWoo Youn$^b$}

\address{$^a$ Korea University, $^b$ University of Maryland}

\ead{swcho@fnal.gov}

\begin{abstract}
 We report a measurement of the fraction of top quark pair events produced via gluon-gluon fusion
 in $p\bar{p}$ collisions at ~$\sqrt{s} = 1.96 ~\rm TeV$ in lepton+jets final states using the full RunII data set corresponding to $9.7 ~\rm fb^{-1}$ of integrated luminosity collected by the D\O\ experiment. 
We utilize a boosted decision tree to distinguish top quark pair events produced by ~$q\bar{q}$ annihilation and ~$gg$ ~fusion. 
We perform a template fit to extract the $t\bar{t}$ production fraction via $gg$ fusion and find $f_{gg} = 0.096 \pm 0.039 ~(\rm stat.) ~^{+0.077}_{-0.062} ~(\rm syst.)$. 
\end{abstract}

\section{Introduction}
 
 The standard model (SM) predicts that at the Tevatron $t\bar{t}$ events are produced predominantly by either quark-antiquark ($q\bar{q}$) annihilation and $gg$ fusion with fractions of $\approx$ 85\% and $\approx$ 15\%, respectively. However, this prediction for the fraction of $t\bar{t}$ production from $gg$ fusion can vary from 10\% to 20\% due to uncertainties on the parton density functions (PDF) \cite{gg1} \cite{gg2}. 
A precise measurement of this quantity will be helpful to have better understanding of the structure of proton. 
Since different production mechanisms can result in significantly different kinematic properties \cite{gg3}, 
a deviation from the SM calculations may also suggest the possible existence of new physics.

The top quark, with a mass of about $175$ $\rm GeV/c^{2}$, has a life time that is an order of magnitude smaller than 
the typical quantum chromodynamics (QCD) hadronization time of $\approx$ $5 \times10^{-24}$ \rm s \cite{top life time}.
As a consequence, the spin information of the top quark is preserved to its decay particles. 
Due to different spin structures for different production modes, the angular distributions of decay particles are useful to distinguish between $q\bar{q}$ annihilation and $gg$ fusion events. 
Furthermore, the gluon has a large degrees of freedom in color charge state than a quark does, the former is involved with the QCD effects such as initial/final state radiations more than the latter, resulting in, 
for example, extra partons and thus affecting the kinematic distributions of the final state particles.

To separate $t\bar{t}$ events produced through $q\bar{q}$ annihilation and $gg$ fusion, we utilize the boosted decision tree (BDT) \cite{BDT}
in the toolkit for multivariate analysis method with ROOT (TMVA) package. 
We use 16 input variables sensitive to the production mechanism, which will be described in Sec. 2. 
We construct the BDT templates from simulated events for $q\bar{q}$ annihilation and $qq$ fusion mechanisms, 
and perform a template fit to measure the fraction of $t\bar{t}$ events produced by $gg$ fusion ($f_{gg}$).

\section{Samples and Event Selection}

\subsection{Data}
This analysis is based on a total integrated luminosity of $9.7 ~\rm fb^{-1}$  collected with the D\O\ detector 
between April 2002 and September 2011 (the full D\O\ RunII data set).

\subsection{Monte Carlo Samples}

\subsubsection{Signal Sample}

We use {\tt MC@NLO} \cite{MCNLO} event generator with {\tt CTEQ6.1M} \cite{cteq} PDF set for $p\bar{p}$ $\rightarrow$ $t\bar{t}$ signal modeling, at $\sqrt s = 1.96$ $\rm TeV$.
We assume a top quark mass of 172.5 \rm GeV and SM $t\bar{t}$ ~spin correlation.
We use {\tt HERWIG} \cite{HERWIG} to simulate the fragmentation of the partons in the decay of top quark, and {\tt GEANT} \cite{GEANT} is followed for the full detector simulation.
We overlay events from random beam crossing on the MC events to model the effect of detector noise and additional $p\bar{p}$ interactions. 
The signal sample is normalized to the cross section of 7.48 \rm pb, which is suggested by the next-to-leading order (NLO) calculation \cite{NLO}.  

\subsubsection{Physical Background Samples}

We consider the following physical background processes in this analysis : {\tt $t\bar{t}$ production, dilepton channels}, {\tt Single top production}, {\tt Di-boson production}, {\tt $Z$+jets production}, {\tt $W$+jets production}.

\subsubsection{Multijet Background Sample}\label{sec:multijet}

In the $e$+jets channel, a jet with high electromagnetic fraction can mimic an electron, and for the $\mu$+jets channel, a muon decaying from heavy flavor particle may appear isolated.
We use a Matrix Method to estimate the contribution of such background events due to lepton misidentification from data.

\subsection{Lepton+jets Event Selection}

The following major selection criteria are applied :

\begin{itemize}
\item {\tt SuperOR} trigger selection for both $e$+jets and  $\mu$+jets channel (single muon OR for $\mu$+jets channel in RunIIa).
\end{itemize}

\begin{itemize}
\item Primary vertex cut : $|z_{PV}| < $ 60 \rm cm.
\end{itemize}

\begin{itemize}
\item Exactly one isolated electron of {\tt emvPoint1} quality or muon of {\tt mediumnseg3} quality which has matched charged track quality of {\tt track new medium} and {\tt TopScaledTight} isolation 
with $p_{T} > $ 20 \rm GeV, $|\eta| < $ 1.1 for electron and $|\eta| < $ 2.1 for muon, and electrons are vetoed in $\mu$+jets channel and muons are vetoed in $e$+jet channel.
\end{itemize}

\begin{itemize}
\item At least four jets with $p_{T} > $ 20 \rm GeV, the leading jet $p_{T} > $ 40 \rm GeV, $|\eta| < $ 2.5, and vertex confirmation is required for RunIIb.
\end{itemize}

\begin{itemize}
\item At least two b-tagged jets based on a tagging algorithm, {\tt MVA BL} tagger, with {\tt L4} working point corresponding to {\tt MVA} $ > $ 0.035.
\end{itemize}

\begin{itemize}
\item Missing transverse energy $ > $ 20 \rm GeV.
\end{itemize}

\subsection{$t\bar{t}$ Reconstruction}

For the $t\bar{t}$ event reconstruction, we find the solution which makes 
the absolute mass difference between the leptonic ($t_{l}$, leptonic $W$ + $b$-jet) and the hadronic ($t_{h}$, hadronic $W$ + $b$-jet) tops ($dm_{t\bar{t}}=|m_{t_{h}} - m_{t_{l}}|$) minimal.

\section{gg Fraction Measurement}

\subsection{Inputs to BDT}\label{sec:input}
We use 16 input variables, which are classified into three categories : 4 kinematic, 9 top-spin related, and 3 topological variables, for the BDT training.

\subsubsection{Top Kinematic Variables}\label{sec:kine}
The four top quark kinematic variables are : {$\Delta\phi_{t\bar{t}} = \phi_{t_{h}} - \phi_{t_{l}}$}, {\tt njt}, {$\beta_{t_{l}}$}, {$\beta_{t_{h}}$}.

\subsubsection{Top-Spin Related Variables}\label{sec:topspinvar}
And we have nine variables related to the top quark spin : {$\Delta\phi_{lhq} = \phi_{lepton} - \phi_{harder ~quark}$}, {$\Delta\phi_{lsq} = \phi_{lepton} - \phi_{softer ~quark}$}, {cos $\theta^{*}_{top}$}, {cos $\theta^{*}_{Wh}$}, {cos $\theta^{*}_{hq}$}, {cos $\theta^{*}_{sq}$}, {cos $\theta^{*}_{Wl}$}, {cos $\theta^{*}_{lep}$}, {cos $\theta^{*}_{\nu}$}.

\subsubsection{Topological Variables}\label{sec:topspin}
We use three topological variables : {\tt Aplanarity}, {\tt Sphericity}, {\tt Centrality}.

\subsection{BDT Templates}
To obtain a single quantity discriminating between $t\bar{t}$ events produced through $q\bar{q}$ annihilation and $gg$ fusion, the sixteen variables are fed into BDT.
We train the BDT on the $q\bar{q}$ annihilation (as the background) and $gg$ fusion (as the signal) distributions of $t\bar{t}$ signal MC events to get the {\tt BDT weight}, using TMVA.
Then, this {\tt BDT weight} is applied to the $q\bar{q}$ annihilation and $gg$ fusion of the $t\bar{t}$ signal, background, and data events 
to get the BDT output distributions (BDT templates).

\subsection{Template Fit}\label{sec:fitting}
To measure the $gg$ fraction, we fit the $gg$ and $q\bar{q}$ templates to the BDT distribution of background subtracted data ({\tt Background Subtracted Data BDT template}).
The template fit function is defined by 
$n^{i}_{BSdata} = [f_{gg} \times n^{i}_{gg} + (1-f_{gg}) \times n^{i}_{qq}] $, 
where $f_{gg}$ is the $gg$ fraction, $n^{i}_{BSdata}$ is the bin contents of the $i_{th}$ bin in the background subtracted data template and
$n^{i}_{gg/qq}$ is the bin contents of the $i_{th}$ bin in the $gg$/($q\bar{q}$) template.
In this template fit, we normalize $gg$ and $q\bar{q}$ templates to the number of background subtracted data.

\begin{table}[h]
\caption{\label{label}The template fit results for $e$, $\mu$, and $l$+jets channels.}
\lineup
\begin{center}
\begin{tabular}{lll}
\br
Template Fit Results & \m\m{gg} Fraction & Goodness of fit\\
\mr
\m $e${\tt +jets channel} & \m $0.077 \pm 0.055$ {\tt(stat.)} & \m\m {\tt 1.04}\\
\m $\mu${\tt +jets channel} & \m $0.116 \pm 0.057$ {\tt(stat.)} & \m\m {\tt 0.94}\\
\m $l${\tt +jets channel} & \m $0.096 \pm 0.039$ {\tt(stat.)} & \m\m {\tt 1.13}\\
\br
\end{tabular}
\end{center}
\end{table}

\subsection{Ensemble Test}\label{sec:ensemble}
For each time of template fit, we make an ensemble test to calibrate fitter and to estimate the expected sensitivity and uncertainty of the measurement.
A single ensemble consists of events randomly selected (for the number of background subtracted data sample) from the signal MC ($gg$ and $q\bar{q}$) templates according to the sample composition.
We consider 5 different signal composition (mixing fractions) between $t\bar{t}$ events from $gg$ fusion and $q\bar{q}$ annihilation with a step of 25 \%.
In example, $f_{gg}$ : $f_{qq}$ = {\tt 0.0:1.0, 0.25:0.75, 0.50:0.50, 0.75:0.25, 1.0:0.0}.
Signal events are randomly selected from the corresponding signal templates according to the given fraction. For each mixing fraction point, 1,000 ensembles are sampled, and the same fitting procedure as in Sec. \ref{sec:fitting} is repeated per ensemble.

\section{Systematic Uncertainties}\label{sec:sys}

Sources of systematic uncertainties are classified into two categories : one affecting the normalization, the other changing shape of the BDT templates.

\subsection{Normalization Systematics}\label{sec:normsys}
This systematic category includes integrated luminosity, trigger requirement, lepton identification, normalization of background, and the determination of multijet background. 
Each systematic source, we use the templates with the systematic effects and repeat the template fit to background subtracted data template distribution.
The deviation of the resulting values from the those obtained using the nominal (central) templates are taken as the corresponding normalization systematic uncertainties.

\subsection{Template Shape Changing Systematics}\label{sec:shapesys}
Template shape changing systematic sources are jet energy scale,  jet energy resolution, sample dependent (jet response) correction, jet identification, vertex confirmation, b identification, PDF, and choice of top quark mass.
In this systematic study, we re-build the templates with the systematic effect reflected for each systematic source by applying the {\tt BDT weight} from the nominal {\tt MC@NLO+HERWIG} sample. 
And we repeat the templates fit to the background subtracted data template distribution.

\section{Result}

From the template fit, we measure the fraction of $t\bar{t}$ production via gg fusion as $f_{gg} = 0.096 \pm 0.039 ~(\rm stat.) ~^{+0.077}_{-0.062} ~(\rm syst.)$ for the top quark mass of 172.5 \rm GeV.
This is good agreement with the standard model prediction of $f_{gg} = 0.150 \pm 0.05$ \cite{gg1}  \cite{gg2}.

\section{Conclusion}

In conclusion, we report a measurement of the fraction of top quark pair production via gluon-gluon fusion in $p\bar{p}$ collisions atÊ~$\sqrt{s} = 1.96 ~\rm TeV$ 
in lepton+jets final states using the full RunII data corresponding to $9.7 ~\rm fb^{-1}$ of integrated luminosity collected by the D\O\ experiment. 
WeÊ~utilize the boosted decision tree to distinguish top quark pair events initiated byÊ~$q\bar{q}$ annihilation andÊ~$gg$Ê~fusion. 
We perform a template fit to extract the $t\bar{t}$ production fraction via $gg$ fusion and find $f_{gg} = 0.096 \pm 0.039 ~(\rm stat.) ~^{+0.077}_{-0.062} ~(\rm syst.)$.

\end{document}